\pdfoutput=1
\RequirePackage{ifpdf}
\ifpdf 
\documentclass[pdftex]{sigma}
\else
\documentclass{sigma}
\fi

\usepackage{cancel}
\numberwithin{equation}{section}

\newtheorem{Proposition}{Proposition}[section]

\renewcommand\vec[1]{\boldsymbol{#1}}
\renewcommand\o{\xi}
\renewcommand\i{\eta}
\newcommand\myvar[2]{\ifcase#2 ? \or #1 \or \bar{#1} \else {#1_{#2}} \fi}
\newcommand\A[1]{\myvar{A}{#1}}
\renewcommand\c[1]{\myvar{c}{#1}}
\newcommand\C[1]{\myvar{C}{#1}}
\renewcommand\L[1]{\myvar{L}{#1}}
\newcommand\G[1]{\myvar{G}{#1}}

\newcommand\ba[1]{\ifcase#1 ? \or a \or \bar{a} \or b \or \bar{b} \else ? \fi }
\newcommand\ka[1]{\ifcase#1 ? \or \alpha \or \bar{\alpha} \or \beta \or \bar{\beta} \else ? \fi }
\renewcommand\u[1]{\ifcase#1 ? \or u_{1} \or u_{2} \or u \or \bar{u} \else u_{\scriptscriptstyle #1} \fi }
\renewcommand\v[1]{\ifcase#1 ? \or w_{1} \or w_{2} \or v \or \bar{v} \or ? \or ? \or w_{3} \or w_{4} \else ? \fi }
\newcommand\dirac{\cancel{\partial}}
\newcommand\mymat[1]{\mathsf{#1}}
\newcommand\myket[1]{|#1\rangle}
\newcommand\mybra[1]{\langle#1|}
\newcommand\transposed{^{\rm \scriptscriptstyle T}}

\begin{document}

\newcommand{\arXivNumber}{2009.01585}

\renewcommand{\PaperNumber}{144}

\FirstPageHeading

\ShortArticleName{Solitons of Some Nonlinear Sigma-Like Models}

\ArticleName{Solitons of Some Nonlinear Sigma-Like Models}

\Author{V.E.~VEKSLERCHIK}
\AuthorNameForHeading{V.E.~Vekslerchik}

\Address{Usikov Institute for Radiophysics and Electronics, \\
 12 Proskura Str., Kharkiv, 61085, Ukraine}
\Email{\href{mailto:vekslerchik@yahoo.com}{vekslerchik@yahoo.com}}

\ArticleDates{Received September 04, 2020, in final form November 30, 2020; Published online December 25, 2020}

\Abstract{We present a set of differential identities for some class of matrices. These identities are used to derive the $N$-soliton solutions for the Pohlmeyer nonlinear sigma-model, two-dimensional self-dual Yang--Mills equations and some modification of the vector Calapso equation.}


\Classification{37J35; 11C20; 35C08; 15B05}

\section{Introduction}

This paper is a continuation of a cycle of works devoted to the derivation of
the soliton solutions for various integrable models
(see \cite{V14,V15,V17,V19c,V19a,V19b} and references therein).
In all these studies we exploit the already known fact that soliton solutions
of almost all integrable equations possess similar structure
which can be clearly expressed in terms of some class of matrices~\cite{V14,V15}.

In this work we consider the following three models. The first one is the Pohlmeyer nonlinear sigma-model \cite{BG87,G77,G79,L77, P76,PR79}, described by the action
\begin{equation}\label{mod:sigma}
 \mathcal{S} = \int {\rm d}\o\, {\rm d}\i \left\{ \frac{ (\partial_{\o}\u3) (\partial_{\i}\u4) } { 1 - \u3 \u4 }
 \pm \u3\u4 \right\},
\end{equation}
where $\partial_{\zeta}$ stands for $\partial / \partial\zeta$, which, depending on whether the variables are real or complex and
on the choice of the involution $\u4 = \kappa \u3^{*}$ (with the asterix indicating the complex conjugation and $\kappa^{2}=1$)
becomes either the ${\rm O}(4)$ sigma-model studied in \cite{G77,L77,PR79} or the ${\rm O}(3,1)$ sigma-model discussed in~\cite{V94,V96}.

Also we consider a modification of the vector Calapso equation
\begin{equation}\label{mod:calapso}
 \dirac^{2} \vec\varphi = f \vec\varphi,
\end{equation}
where $\vec\varphi$ is a complex 4-vector, $\vec\varphi \in \mathbb{C}^{4}$, $\dirac$ is a two-dimensional Dirac-type operator $\dirac = \gamma^{\xi}\partial_{\xi}+\gamma^{\eta}\partial_{\eta}$, where $\gamma^{\xi}$, $\gamma^{\eta}$ are two Dirac matrices and $f$ is some function of $\dirac\vec\varphi$ (we will specify both $\dirac$ and $f$ in Section~\ref{sec:calapso}).

The third equation discussed in this paper,
\begin{equation}\label{mod:sdym}
 \partial_{\o} \big( \mymat{U}^{-1} \partial_{\i} \mymat{U} \big)
 + \big[ \mymat{\sigma}_{3}, \mymat{U}^{-1} [ \mymat{\sigma}_{3}, \mymat{U} ] \big] = 0,
\end{equation}
where $\mymat{U}$ is a $2 \times 2$ complex matrix, can be considered as a two-dimensional reduction of the self-dual Yang--Mills equations~\cite{MZ81,P80, Y77}.

As in the works cited above, we do not address in this paper the questions of
integrability. The models \eqref{mod:sigma} and \eqref{mod:sdym} are already
known to be integrable. We are not aware of any discussion in the literature
of the particular version \eqref{mod:calapso} of the Calapso equation.
However, its relationships with other integrable models strongly indicate that
is integrable as well. Moreover, in what follows we present the $N$-soliton
solutions for \eqref{mod:calapso} which means that it passes the so-called
`three-soliton' test (see, for example \cite{H87, R81} or Section~8.2 of the
book~\cite{HJN16}) which is another evidence of its integrability.
The main task of this work is to obtain the $N$-soliton solutions for the models
\eqref{mod:sigma}--\eqref{mod:sdym}, and to accomplish it we use some kind of the
`direct' method which can be viewed as a generalization or modification of the
Cauchy matrix approach \cite{FN17,HJN16,NAH09, NQC83,QNCL84,ZZ13}.
This method is based on using some class of matrices, the so-called
`almost-intertwining' matrices~\cite{KG01} that
satisfy the `rank one condition' \cite{GK02,GK06b,GK06a},
which is a~particular case of the Sylvester equation \cite{DM10a, DM10b, SZN17, XZZ14}.
The key result of the cited works as well as of many other studies
(see, e.g., \cite{CS11, S93,S01,S10}) is the fact that these matrices clearly
describe structure of the soliton solutions for a large number of integrable
models.

We introduce in Section~\ref{sec:aux} some functions constructed of the soliton
matrices studied in \cite{V15} and present a set of algebraic and differential
identities for these functions (we will use the term `auxiliary system' for
this set and hope that a reader will not confuse it with the auxiliary systems
used in the framework of the inverse scattering transform).
Then we demonstrate in Section~\ref{sec:solitons}, by some elementary calculations, that functions
satisfying this auxiliary system can be used to construct solutions for the
equations that we study in this paper. In Section~\ref{sec:involution} we focus
on the questions related to the complex conjugation and show that to
perform the corresponding reductions in the framework of our direct approach is
much easier than in the framework of the inverse scattering transform,
the dressing method or the algebro-geometric approach. Finally, in the last section
we give a few comments about the obtained results.

\section{Auxiliary system}\label{sec:aux}

We start with the matrices defined by the equation
\begin{equation}
 \mymat{\L2} \mymat{\A1} - \mymat{\A1} \mymat{\L1} = \myket{\ka1} \mybra{\ba1},
\qquad
 \mymat{\L1} \mymat{\A2} - \mymat{\A2} \mymat{\L2} = \myket{\ka2} \mybra{\ba2}.\label{eq:sy}
\end{equation}
Here,
$\mymat{\L1}$ and $\mymat{\L2}$ are diagonal constant $N \times N$ matrices,
$\mymat{\L1} = \operatorname{diag} ( \L1_{1}, {\dots}, \L1_{N} )$,
$\mymat{\L2} = \operatorname{diag} ( \L2_{1}, {\dots}, \L2_{N} )$,
$\myket{\ka1}$ and $\myket{\ka2}$ are constant $N$-component columns,
$\myket{\ka1} = ( \myvar\alpha1_{1}, \dots, \myvar\alpha1_{N})\transposed$,
$\myket{\ka2} = ( \myvar\alpha2_{1}, \dots, \myvar\alpha2_{N})\transposed$,
and~$\mybra{\ba1}$ and~$\mybra{\ba2}$ are $N$-component rows that depend on the
coordinates describing the model.

It should be noted that throughout this paper the overbar \emph{does not} mean
the complex conjugation (which will be indicated by the asterisk).

The $\o$- and $\i$-dependence of the matrices $\mymat{\A1}$ and $\mymat{\A2}$
that we use in this work is defined by
\begin{equation}\label{def:evol-a}
 \langle\ba1(\o,\i)| = \mybra{\c1} \exp\big( \o \mymat{\L1}^{-1} - \i \mymat{\L1} \big), \qquad
 \langle\ba2(\o,\i)| = \mybra{\c2} \exp\big( {-} \o \mymat{\L2}^{-1} + \i \mymat{\L2} \big),
\end{equation}
where
$\mybra{\c1} = (\c1_{1}, \dots, \c1_{N} )$
and
$\mybra{\c2} = (\c2_{1}, \dots, \c2_{N} )$
are arbitrary constant rows,
which implies, together with~\eqref{eq:sy},
\begin{equation}
 \mymat{\A1}(\o,\i) = \mymat{\C1} \exp\big( \o \mymat{\L1}^{-1} - \i \mymat{\L1} \big),\qquad
 \mymat{\A2}(\o,\i) = \mymat{\C2} \exp\big( {-} \o \mymat{\L2}^{-1} + \i \mymat{\L2} \big),
\label{def:evol}
\end{equation}
where $\mymat{\C1}$ and $\mymat{\C2}$ are constant $N \times N$ matrices given by
\begin{equation*}
 \mymat{\C1} = \left( \frac{ \myvar\alpha1_{j} \c1_{k} }{ \L2_{j} - \L1_{k} } \right)_{j,k=1}^{N},
 \qquad
 \mymat{\C2} = \left( \frac{ \myvar\alpha2_{j} \c2_{k} }{ \L1_{j} - \L2_{k} } \right)_{j,k=1}^{N}.
\end{equation*}
Note that $\mymat{\A1}$ (as well as $\mymat{\A2}$) satisfies
$\partial_{\xi\eta} \mymat{\A1} + \mymat{\A1} = 0$,
an equation that naturally appears as a~linearization of, say, the sigma model
discussed in this paper. In the framework of the inverse scattering transform,
the exponential functions like ones in~\eqref{def:evol-a} and~\eqref{def:evol}
arise as solutions of the linear equations governing evolution of the
scattering data and, again, they usually have the form of the dispersion
relations of the corresponding linearized problems.

However, in the context of this study equations~\eqref{def:evol-a}
or~\eqref{def:evol} can be viewed as the \emph{definitions} that constitute a part of
the proposed \emph{ansatz}.

Now, our task is to calculate derivatives of various combinations of
the matrices $\mymat{\A1}$ and $\mymat{\A2}$,
matrices $\mymat{\G1}$ and $\mymat{\G2}$ defined by
\begin{equation}\label{def:G}
 \mymat{\G1} = \big( 1 + \mymat{\A1} \mymat{\A2} \big)^{-1},
\qquad
 \mymat{\G2} = \big( 1 + \mymat{\A2} \mymat{\A1} \big)^{-1},
\end{equation}
rows $\mybra{\ba1}$, $\mybra{\ba2}$ and
columns $\myket{\ka1}$, $\myket{\ka2}$.
In particular, we are going to derive a closed set of differential identities
involving the eight functions
\begin{alignat}{3}
& \u3 = 1 - \mybra{ \ba2} \mymat{\G1} \mymat{\A1} \myket{ \ka4}, \qquad && \v3 = \mybra{ \ba2} \mymat{\G1} \myket{ \ka3},&\nonumber\\
& \u4 = 1 - \mybra{ \ba1} \mymat{\G2} \mymat{\A2} \myket{ \ka3}, \qquad && \v4 = \mybra{ \ba1} \mymat{\G2} \myket{ \ka4} & \label{def:u3v4}
\end{alignat}
and
\begin{alignat}{3}
& \v1 = \mybra{ \ba2} \mymat{\G1} \myket{ \ka1}, \qquad && \v7 = \mybra{ \ba4} \mymat{\G1} \myket{ \ka3},& \nonumber\\
& \v2 = \mybra{ \ba1} \mymat{\G2} \myket{ \ka2},\qquad && \v8 = \mybra{ \ba3} \mymat{\G2} \myket{ \ka4},&\label{def:v1v8}
\end{alignat}
where
\begin{alignat}{3}
& \mybra{ \ba3} = \mybra{ \ba1} \mymat{\L1}^{-1}, \qquad && \myket{ \ka3} = \mymat{\L2}^{-1} \myket{ \ka1},&\nonumber\\
& \mybra{ \ba4} = \mybra{ \ba2} \mymat{\L2}^{-1},\qquad && \myket{ \ka4} = \mymat{\L1}^{-1} \myket{ \ka2}.& \label{def:baka}
\end{alignat}

Using straightforward calculations one can obtain the following identities involving the $\partial_{\o}$-derivatives.

\begin{Proposition} \label{prop:aux-o}
Functions $\u3$, $\u4$, $\v3$, $\v3$, $\v4$, $\v1$, $\v2$
defined in \eqref{def:u3v4} and \eqref{def:v1v8}
satisfy the following set of equations
\begin{alignat}{4}
& \partial_{\o} \u3 = - \v3 \v8,\qquad && \partial_{\o} \v3 = - \u3 \v7,\qquad && \partial_{\o} \v1 = - \u3 \v3,&\nonumber\\
& \partial_{\o} \u4 = \v4 \v7,\qquad && \partial_{\o} \v4 = \u4 \v8,\qquad && \partial_{\o} \v2 = \u4 \v4.& \label{do:all}
\end{alignat}
\end{Proposition}

In a similar way, one can derive the set of $\partial_{\i}$-identities.

\begin{Proposition} \label{prop:aux-i}
Functions $\u3$, $\u4$, $\v3$, $\v3$, $\v4$, $\v7$, $\v8$ defined in \eqref{def:u3v4} and \eqref{def:v1v8} satisfy the following set of equations
\begin{alignat}{4}
& \partial_{\i} \u3 = - \v4\v1,\qquad && \partial_{\i} \v3 = \u4 \v1, \qquad && \partial_{\i} \v7 = \u4 \v3,&\nonumber\\
& \partial_{\i} \u4 = \v3\v2,\qquad && \partial_{\i} \v4 = - \u3 \v2,\qquad &&\partial_{\i} \v8 = - \u3 \v4. &\label{di:all}
\end{alignat}
\end{Proposition}

We do not present here a proof of all of these identities. In Appendix~\ref{app:ids} a reader can find examples of how to obtain some
of them, while the rest can be derived in an analogous way.

An immediate consequence of these results is that the function $I$ defined as $I = \u3 \u4 + \v3 \v4$ is constant: $\partial_{\o}I = \partial_{\i}I = 0$. More careful analysis leads to the identity
\begin{equation}\label{unity}
 \u3 \u4 + \v3 \v4 = 1,
\end{equation}
which will be often used in what follows. It turns out that a derivation of this simple identity is the
most cumbersome part of the calculations of this paper. We present a proof of \eqref{unity} in Appendix~\ref{app:unity}.

System \eqref{do:all}--\eqref{unity} is not new. It is closely related to the Ablowitz--Ladik hierarchy~\cite{AL75},
which is not surprising because, as is shown in~\cite{V15}, the bright solitons of the Ablowitz--Ladik hierarchy are built of matrices $\mymat{\A1}$ and $\mymat{\A2}$ \eqref{eq:sy} and have the structure of functions defined in~\eqref{def:u3v4}. However, we do not discuss these questions here and consider \eqref{do:all}--\eqref{unity} as a closed set of identities which is used in what follows to construct explicit solutions for the equations which
are the subject of this paper.

\section[N-soliton solutions]{$\boldsymbol{N}$-soliton solutions}\label{sec:solitons}

In this section we construct $N$-soliton solutions for the equations listed in the introduction. Identities \eqref{do:all}--\eqref{unity} give us a possibility to do this with very little effort.

\subsection{Nonlinear sigma-model}\label{sec:sigma}

Starting from equations \eqref{do:all} and \eqref{di:all}, it is easy to derive the following identities involving second derivatives of the functions $\u3$ and $\u4$
\begin{gather}
 \v3 \v4 \partial_{\o\i} \u3 + \u4 (\partial_{\o} \u3 ) (\partial_{\i} \u3 ) = \u3 \v3^{2} \v4^{2},\label{osm:u3}\\
 \v3 \v4 \partial_{\o\i} \u4 + \u3 (\partial_{\o} \u4 ) (\partial_{\i} \u4 ) = \u4 \v3^{2} \v4^{2}.\label{osm:u4}
\end{gather}
Noting that $\v3\v4 = 1 - \u3\u4$, we can rewrite equations~\eqref{osm:u3} and \eqref{osm:u4} as
\begin{gather*}
 U^{-1} \partial_{\o\i} \u3 + U^{-2}(\partial_{\o} \u3) (\partial_{\i} \u3) \u4 = \u3,\\
 U^{-1} \partial_{\o\i} \u4 + U^{-2} \u3 (\partial_{\o} \u4) (\partial_{\i} \u4) = \u4,
\end{gather*}
where
\begin{equation*}
 U = 1 - \u3 \u4.
\end{equation*}
These equations are nothing but the Euler equations
\begin{equation*}
 \delta \mathcal{S} / \delta\u3 =
 \delta \mathcal{S} / \delta\u4 = 0
\end{equation*}
for the action
\begin{equation*}
 \mathcal{S} = \int {\rm d}\o {\rm d}\i \, \mathcal{L}
\end{equation*}
with the Lagrangian given by
\begin{equation*}
 \mathcal{L} = \frac{(\partial_{\o} \u3)(\partial_{\i} \u4) } { 1 - \u3\u4} + \u3\u4.
\end{equation*}

In a similar way, one can derive from \eqref{do:all} and \eqref{di:all} the
identities
\begin{gather*}
 \u3 \u4 \partial_{\o\i} \v3 + (\partial_{\o} \v3) (\partial_{\i} \v3) \v4 = - \u3^{2} \u4^{2} \v3,\\
 \u3 \u4 \partial_{\o\i} \v4 + \v3 (\partial_{\o} \v4) (\partial_{\i} \v4) = - \u3^{2} \u4^{2} \v4,
\end{gather*}
rewrite them as
\begin{gather*}
 V^{-1} \partial_{\o\i} \v3 + V^{-2} (\partial_{\o} \v3) (\partial_{\i} \v3) \v4 = - \v3,\\
 V^{-1} \partial_{\o\i} \v4 + V^{-2} \v3 (\partial_{\o} \v4) (\partial_{\i} \v4) = - \v4,
\end{gather*}
where
\begin{equation*}
 V = 1 - \v3 \v4,
\end{equation*}
and note that they correspond to the Lagrangian
\begin{equation*}
 \mathcal{L} = \frac{(\partial_{\o} \v3)(\partial_{\i} \v4) } { 1 - \v3\v4 } - \v3\v4.
\end{equation*}

These calculations can be summarized as follows.

\begin{Proposition} \label{prop:sigma}
Functions $\u3$, $\u4$, $\v3$, $\v3$ and $\v4$ defined in~\eqref{def:u3v4} satisfy the Euler equations for the Lagrangian
\begin{equation}
 \mathcal{L} = \frac{(\partial_{\o} w )(\partial_{\i} \bar{w}) } { 1 - w \bar{w} } \pm w \bar{w}.\label{eq:sigma}
\end{equation}
\end{Proposition}

Thus, functions defined in \eqref{def:u3v4} provide solutions for the field equations for the Pohlmeyer nonlinear sigma-model.

\subsection{Calapso equation}\label{sec:calapso}

The elementary consequence of the equations \eqref{do:all} and \eqref{di:all} is the fact that the four-vector $\vec{\varphi}$ defined by
\begin{equation}
 \vec{\varphi} = ( \v8, \v1, \v7, \v2 )\transposed\label{cae:varphi}
\end{equation}
obeys the identity
\begin{equation}
 \partial_{\o\i} \vec{\varphi} = f \vec{\varphi},\label{cae:ddv}
\end{equation}
where
\begin{equation*}
 f = \v3 \v4 - \u3 \u4.
\end{equation*}
Moreover, one can express the function $f$ in terms of $\v1$, $\v2$, $\v7$ and $\v8$ by noting that
\begin{equation*}
(\partial_{\o} \v1) (\partial_{\o} \v2) + (\partial_{\i} \v7) (\partial_{\i} \v8) = - 2 \u3\u4\v3\v4,
\end{equation*}
which, together with \eqref{unity}, leads to
\begin{equation}
(\partial_{\o} \v1) (\partial_{\o} \v2) + (\partial_{\i} \v7) (\partial_{\i} \v8) = \tfrac{1}{2} \big( f^{2} - 1 \big).\label{cae:f}
\end{equation}
This means that \eqref{cae:ddv} can be presented as a closed equation for the vector $\vec\varphi$.

After introducing the Dirac operator by
\begin{equation*}
 \dirac = \left(\begin{matrix}
 0 & 0 & 0 & \partial_{\o} \\
 0 & 0 & \partial_{\i} & 0 \\
 0 & - \partial_{\o} & 0 & 0 \\
 - \partial_{\i} & 0 & 0 & 0
 \end{matrix}\right),
\end{equation*}
noting that $\partial_{\o\i} = - \dirac^{2}$ and rewriting the identity \eqref{cae:f} in terms of $\vec\varphi$ as
\begin{equation*}
 f^{2} = 1 - (\dirac\vec\varphi)\transposed \mymat{\gamma}^{5} (\dirac\vec\varphi),
\end{equation*}
where
\begin{equation*}
 \mymat{\gamma}^{5} =
 \left(\begin{matrix}
 0 & 0 & 1 & 0 \\
 0 & 0 & 0 & 1 \\
 1 & 0 & 0 & 0 \\
 0 & 1 & 0 & 0
 \end{matrix}\right),
\end{equation*}
one can present this result as

\begin{Proposition} \label{prop:calapso}
Vector $\vec\varphi$ given by \eqref{cae:varphi} where functions $\v1$, $\v2$, $\v7$ and $\v8$ are defined in~\eqref{def:v1v8}, is a solution for the equations
\begin{equation}
 \dirac^{2} \vec\varphi = f \vec\varphi\label{eq:calapso}
\end{equation}
with
\begin{equation}
 f = \sqrt{ 1 - (\dirac\vec\varphi)\transposed \mymat{\gamma}^{5} (\dirac\vec\varphi) }.\label{eq:calapso-f}
\end{equation}
\end{Proposition}

Rewriting this statement in terms of the vector $\vec\phi$,
\begin{equation*}
 \vec\phi = \dirac\vec\varphi,
\end{equation*}
one can easily obtain the following corollary

\begin{Proposition} \label{prop:vsigma}
Vector $\vec\phi$ given by
\begin{equation*}
 \vec\phi = ( \u4\v4, \u4\v3, \u3\v3, \u3\v4 )\transposed,
\end{equation*}
where functions $\u3$, $\u4$, $\v3$ and $\v4$ are defined in \eqref{def:u3v4} satisfies equation
\begin{equation*}
 \dirac \frac{ \dirac\vec\phi } { \sqrt{ 1 - {\vec\phi}\transposed \mymat{\gamma}^{5} \vec\phi } } = \vec\phi,
\end{equation*}
which describes some vector variant of the sigma model discussed above.
\end{Proposition}

\subsection{2D Self-dual Yang--Mills-like equations}

To derive the soliton solutions for the two-dimensional self-dual ${\rm SU}(2)$ Yang--Mills-like equations consider the matrix
\begin{equation}\label{def:U}
 \mymat{U} = \left(\begin{matrix}
 \u3 & - \v4 \\ \v3 & \u4
 \end{matrix}\right),
\end{equation}
which, due to \eqref{unity}, belongs to ${\rm SL}(2,\mathbb{C})$, $\det\mymat{U} = 1$. Equations \eqref{do:all} and \eqref{di:all} imply
\begin{equation*}
 \partial_{\o} \mymat{U} =
 \left(\begin{matrix}
 - \v3 \v8 & - \u4 \v8 \\ - \u3 \v7 & \v4 \v7
 \end{matrix}\right), \qquad
 \partial_{\i} \mymat{U} =
 \left(\begin{matrix}
 - \v4 \v1 & \u3 \v2 \\ \u4 \v1 & \v3 \v2
 \end{matrix}\right).
\end{equation*}
After differentiating these identities once more and using, again, \eqref{do:all} and \eqref{di:all}, one can arrive, after some simple
calculations, at
\begin{equation*}
 \partial_{\o\i} \mathsf{U} -( \partial_{\o} \mathsf{U}) \mymat{U}^{-1} ( \partial_{\i} \mathsf{U} )
= \left(\begin{matrix}
 \u3\v3\v4 & + \u3\u4\v4 \\ -\u3\u4\v3 & \u4\v3\v4
 \end{matrix}\right)
\end{equation*}
and
\begin{equation*}
 \mymat{U}^{-1} (\partial_{\o\i} \mymat{U}) - \mymat{U}^{-1} (\partial_{\o} \mymat{U}) \mymat{U}^{-1} (\partial_{\i} \mymat{U})
 = \left(\begin{matrix}
 0 & \u4\v4 \\ -\u3\v3 & 0
 \end{matrix}\right).
\end{equation*}
It easy to demonstrate that the right-hand side of the last equation can be rewritten as
\begin{equation*}
 \left(\begin{matrix} 0 & \u4\v4 \\ -\u3\v3 & 0 \end{matrix}\right)
 = -\tfrac{1}{4} \big[ \mymat{\sigma}_{3}, \mymat{U}^{-1} [ \mymat{\sigma}_{3}, \mymat{U} ] \big].
\end{equation*}
To summarize, we have derived the following result.

\begin{Proposition} \label{prop:sdYM}The matrix \eqref{def:U} with the functions $\u3$, $\u4$, $\v3$ and $\v4$ defined in \eqref{def:u3v4} satisfies the two-dimensional self-dual Yang--Mills-like equations
\begin{equation}
 \partial_{\o} \big( \mymat{U}^{-1} \partial_{\i} \mymat{U} \big)
 +\tfrac{1}{4} \big[ \mymat{\sigma}_{3}, \mymat{U}^{-1} [ \mymat{\sigma}_{3}, \mymat{U} ] \big] = 0.\label{eq:sdYM}
\end{equation}
\end{Proposition}

It is easy to see that if one starts from the system \eqref{do:all}--\eqref{unity}, then to derive soliton solutions for equations
\eqref{eq:sigma}, \eqref{eq:calapso} with \eqref{eq:calapso-f} or \eqref{eq:sdYM}
becomes a rather easy task.

\section{Solitons with involution} \label{sec:involution}

Till now, we have not raised the questions related to the complex or Hermitian conjugation: the functions $\u3$ and $\u4$, $\v3$ and $\v4$ as well as $\v1$, $\v2$, $\v7$ and $\v8$ have been treated as independent. However, in physical applications of the models discussed in this paper these questions are very important. Thus, in this section we study the properties of the described in the previous section solutions from this viewpoint.

Our first problem is to determine conditions which ensure the following identity:
\begin{equation}
 \mymat{\A2} = \kappa \mymat{\A1}^{*}, \qquad \kappa = \pm 1,\label{inv:A}
\end{equation}
where $*$ stands for the complex conjugation.

Analyzing the compatibility of equations \eqref{eq:sy} and equations \eqref{def:evol}, which determine the $\o$- and $\i$-dependence,
with the complex conjugation on can distinguish two important cases $\mymat{\L2} = (\mymat{\L1}^{*})^{\pm 1}$. In what follows, we consider these cases separately and see how the involution modifies equations~\eqref{eq:sigma},~\eqref{eq:calapso} with~\eqref{eq:calapso-f}, \eqref{eq:sdYM} and their solitons.

\subsection{`Minkowski' case}

In this case the diagonal matrices $\mymat{\L1}$ and $\mymat{\L2}$, the rows $\mybra{\ba1}$ and $\mybra{\ba2}$ and the columns $\myket{\ka1}$ and $\myket{\ka2}$ are related by
\begin{equation}\label{inv:min}
 \mymat{\L2} = \mymat{\L1}^{*}, \qquad
 \mybra{\ba2} = \mybra{\ba1}^{*}, \qquad
 \myket{\ka2} = \kappa\myket{\ka1}^{*},
\end{equation}
while the variables $\o$ and $\i$ should satisfy conditions $\o=-\o^{*}$ and $\i=-\i^{*}$. After introducing real variables $t$ and $x$ by
\begin{equation*}
 \o = {\rm i} (t+x),
 \qquad
 \i = {\rm i} (t-x)
\end{equation*}
the row $\mybra{\ba1}$ and the matrix $\mymat{\A1}$ can be presented as
\begin{equation}\label{min:a}
 \mybra{\ba1} = \mybra{c}\exp[{\rm i}\mymat{\Theta}(t,x)], \qquad
 \mymat{\A1} = \mymat{C}\exp[{\rm i}\mymat{\Theta}(t,x) ].
\end{equation}
Here $\mybra{c}$ is an arbitrary constant row, $\mymat{C}$ is the constant matrix given by
\begin{equation}
 \mymat{C} =
 \left(\frac{ \alpha_{j} c_{k} }{ L_{j}^{*} - L_{k} }
 \right)_{j,k=1}^{N},
\end{equation}
where $c_{j}$, $\alpha_{j}$ and $L_{j}$ are the components of $\mybra{c}$, $\myket{\ka1}$ and $\mymat{\L1}$,
\begin{equation}
 \mybra{c} = ( c_{1}, \dots, c_{N}), \qquad
 \myket{\ka1} = ( \alpha_{1}, \dots, \alpha_{N} )\transposed, \qquad
 \mymat{\L1} = \operatorname{diag}( L_{1}, \dots, L_{N} ),
\label{def:components}
\end{equation}
which play the role of the constant parameters of the $N$-soliton
solutions,\footnote{Strictly speaking, of $3N$ parameters \eqref{def:components}, only $2N$ are
essential. By simple matrix transformation some of them can be eliminated
by redefining the other. Thus, one can put, say, all $\alpha_{j}=1$.}
and
\begin{equation}
 \mymat{\Theta}(t,x) = t ( \mymat{\L1}^{-1} - \mymat{\L1} ) + x ( \mymat{\L1}^{-1} + \mymat{\L1}).\label{min:T}
\end{equation}
The restrictions \eqref{inv:min} lead to the following relations between the functions which are defined in~\eqref{def:u3v4} and~\eqref{def:v1v8}:
\begin{equation}
 \u4 = \u3^{*}, \qquad \v4 = \kappa\v3^{*}, \qquad
 \v2 = \kappa\v1^{*}, \qquad \v8 = \kappa\v7^{*}.\label{inv:minuv}
\end{equation}

Now we can reformulate some of the results presented in the previous section.

\begin{Proposition} \label{prop:sigma-min}
Functions $\u3$ and $\v3$ defined in \eqref{def:u3v4}
together with \eqref{inv:A}, \eqref{inv:min} and \eqref{min:a}--\eqref{min:T}
satisfy the Euler equations for the Lagrangians
\begin{gather*}
 \mathcal{L}_{\u3} = \frac{(\partial^{\mu}\u3)(\partial_{\mu}\u3^{*}) } { 1 - | \u3 |^{2} } - 4 | \u3 |^{2}, 
\qquad \mathcal{L}_{\v3}
 = \frac{(\partial^{\mu}\v3)(\partial_{\mu}\v3^{*})} { 1 - \kappa | \v3 |^{2} } + 4 | \v3 |^{2}, 
\end{gather*}
correspondingly. Here $\partial_{\mu} = (\partial_{t},\partial_{x})$, $\partial^{\mu} = (\partial_{t},-\partial_{x})$ and summation over $\mu$ is understood.
\end{Proposition}

Considering the Calapso equation (Proposition~\ref{prop:calapso}), it should be noted that due to the symmetry~\eqref{inv:minuv} we can rewrite it as an equation for the $\mathbb{C}^{2}$ vectors $\vec\psi = \frac{1}{\sqrt{2}}( \v1, \v7)\transposed$. After the redefinition of the Dirac operator,
\begin{equation*}
 \dirac =
 \left( \begin{matrix}
 0 & \partial_{t} - \partial_{x} \\
	\partial_{t} + \partial_{x} & 0
 \end{matrix}
 \right)\end{equation*}
one can easily verify that
\begin{equation*}
(\dirac\psi)^{\dagger} (\dirac\psi) = 4 |\u3|^{2} |\v3|^{2},
\end{equation*}
which leads to the following result.

\begin{Proposition} \label{prop:calapso-min}Vector
\begin{equation*}
 \vec\psi = \frac{1}{\sqrt{2}} \left(\begin{matrix} \v1 \\ \v7 \end{matrix}\right),
\end{equation*}
where functions $\v1$ and $\v7$ are defined in \eqref{def:v1v8} together with \eqref{inv:A}, \eqref{inv:min} and \eqref{min:a}--\eqref{min:T} is a~solution for the Calapso-like equation
\begin{equation*}
 \dirac^{2} \vec\psi = 4 f \vec\psi
\end{equation*}
with
\begin{equation*}
 f = \sqrt{ 1 - \kappa (\dirac\psi)^{\dagger}(\dirac\psi)}.
\end{equation*}
\end{Proposition}

\subsection{`Euclidean' case}

In this case the diagonal matrices $\mymat{\L1}$ and $\mymat{\L2}$, the rows $\mybra{\ba1}$ and $\mybra{\ba2}$ and the columns $\myket{\ka1}$ and $\myket{\ka2}$ are related by
\begin{equation}\label{inv:euc}
 \mymat{\L2} = \big(\mymat{\L1}^{-1}\big)^{*},
 \qquad
 \mybra{\ba2} = \big( \mybra{\ba1}\mymat{\L1}^{-1}\big)^{*},
 \qquad
 \myket{\ka2} = - \kappa \mymat{\L1} (\myket{\ka1})^{*}
\end{equation}
and $\i=\o^{*}$. After introducing real variables $x$ and $y$ by
\begin{equation*}
 \o = x + {\rm i}y, \qquad \i = x - {\rm i} y.
\end{equation*}
the row $\mybra{\ba1}$ and the matrix $\mymat{\A1}$ can be presented as
\begin{equation}\label{euc:a}
 \mybra{\ba1} = \mybra{c}\exp[ \mymat{\Theta}(x,y) ], \qquad
 \mymat{\A1} = \mymat{C}\exp[ \mymat{\Theta}(x,y)].
\end{equation}
Here $\mybra{c}$ is an arbitrary constant row,
$\mymat{C}$ is the constant matrix given by
\begin{equation}
 \mymat{C} = \left( \frac{ L_{j}^{*} \alpha_{j} c_{k} }{ 1 - L_{j}^{*} L_{k} } \right)_{j,k=1}^{N},
\end{equation}
where $c_{j}$, $\alpha_{j}$ and $L_{j}$ are defined in~\eqref{def:components} and
\begin{equation}
 \mymat{\Theta}(x,y) = x \big( \mymat{\L1}^{-1} - \mymat{\L1} \big) + {\rm i}y \big( \mymat{\L1}^{-1} + \mymat{\L1} \big).\label{euc:T}
\end{equation}

One can show that in this case the functions $\u3$ and $\u4$ are real,
\begin{equation*}
 \operatorname{Im} \u3 = \operatorname{Im} \u4 = 0,
\end{equation*}
while other functions defined in \eqref{def:u3v4} and \eqref{def:v1v8} are related by
\begin{equation*}
 \v4 = - \kappa\v3^{*}, \qquad
 \v7 = - \kappa\v2^{*}, \qquad
 \v8 = - \kappa\v1^{*}.
\end{equation*}

Now we give a few examples of the `physical' forms of the results presented in Section~\ref{sec:solitons}.

\begin{Proposition} \label{prop:sigma-euc}
Function $\v3$ defined in \eqref{def:u3v4} together with \eqref{inv:A}, \eqref{inv:euc} and \eqref{euc:a}--\eqref{euc:T} satisfies the Euler equations for the Lagrangian
\begin{equation*}
 \mathcal{L} = \frac{( \nabla\v3, \nabla\v3^{*}) } { 1 + \kappa | \v3 |^{2} } - 4 | \v3 |^{2},
\end{equation*}
where $\nabla$ is the gradient operator, $\nabla = (\partial_{x},\partial_{y})\transposed$.
\end{Proposition}

As in the `Minkowski' case, we rewrite the Calapso equation as an equation for the $\mathbb{C}^{2}$ vectors $\vec\psi = \frac{1}{\sqrt{2}} ( \v7, \v1 )\transposed$. Defining the two-dimensional Dirac operator by
\begin{equation*}
 \dirac =
 \left(
 \begin{matrix}
 0 & \partial_{x} - {\rm i} \partial_{y} \\
	\partial_{x} + {\rm i} \partial_{y} & 0
 \end{matrix}
 \right)
\end{equation*}
one can obtain that
$(\dirac\psi)^{\dagger} \sigma^{1} (\dirac\psi) = - 4 \u3\u4 |\v3|^{2}$
which leads to the following result.

\begin{Proposition} \label{prop:calapso-euc}Vector
\begin{equation*}
 \vec\psi = \frac{1}{\sqrt{2}} \left(\begin{matrix} \v7 \\ \v1 \end{matrix}\right)
\end{equation*}
with functions $\v1$ and $\v7$ defined in \eqref{def:v1v8} together with \eqref{inv:A}, \eqref{inv:euc} and \eqref{euc:a}--\eqref{euc:T} is a solution for the Calapso-like equation
\begin{equation*}
 \dirac^{2} \vec\psi = - 4 f \vec\psi,
\end{equation*}
where
\begin{equation*}
 \dirac = \sigma^{1} \partial_{x} + \sigma^{2}\partial_{y},
\end{equation*}
$\sigma^{i}$ are the Pauli matrices and
\begin{equation*}
 f = \sqrt{ 1 - \kappa(\dirac\psi)^{\dagger} \sigma^{1} (\dirac\psi)}.
\end{equation*}
\end{Proposition}

Considering the self-dual Yang--Mills equation (see Proposition~\ref{prop:sdYM}), we restrict ourselves with the $\kappa=1$ case.

\begin{Proposition} \label{prop:sdYM-euc}
The matrix
\begin{equation*}
 \mymat{U} = \left(\begin{matrix}
 \u3 & \v3^{*} \\ \v3 & \u4
 \end{matrix}\right),
\end{equation*}
with the functions $\u3$, $\u4$ and $\v3$ defined in \eqref{def:u3v4}, \eqref{euc:a}--\eqref{euc:T} with $\kappa=1$ and $\mymat{\A2}=\mymat{\A1}^{*}$ is a Hermitian solution,
\begin{equation*}
 \mymat{U}^{\dagger}=\mymat{U}
\end{equation*}
for the equation
\begin{equation*}
\frac{\partial}{\partial\o}\big( \mymat{U}^{-1} \frac{\partial}{\partial{\o^{*}}} \mymat{U} \big)
 +\tfrac{1}{4} \big[ \mymat{\sigma}_{3}, \mymat{U}^{-1} [ \mymat{\sigma}_{3}, \mymat{U} ] \big] = 0.
\end{equation*}
\end{Proposition}

\section{Discussion} \label{sec:discussion}

In this paper we have presented the $N$-soliton solutions for several models that appear in the field theory. Considering the Pohlmeyer sigma-model, our results essentially overlap with ones obtained by Barashenkov and Getmanov in~\cite{BG87} by means of the dressing method. At the same time, the solutions for the reduction~\eqref{mod:sdym} of the self-dual Yang--Mills equations and for the vector Calapso equation~\eqref{mod:calapso} seem to be new.

Bearing in mind possible continuations of this work, we would like to give a~few remarks on the questions related to the topic of this paper.

First, we want to give a comment about the interrelations between the discussed models. As one can easily see, all models considered here are closely related to the auxiliary system~\eqref{do:all},~\eqref{di:all}. At the same time, it has been shown in~\cite{V94} that the sigma-model from Section~\ref{sec:sigma} can be described in terms of the Ablowitz--Ladik hierarchy. A similar result has been obtained in \cite{BG87} where the authors derived soliton solutions for different versions of Pohlmeyer sigma-models starting from a closed system of equations (they called it the `$\mathcal{G}$-system') which is, in fact, the set of the simplest equations of the Ablowitz--Ladik hierarchy. This indicates that both vector Calapso equation~\eqref{eq:calapso},~\eqref{eq:calapso-f} and matrix Yang--Mills-type equation~\eqref{eq:sdYM} can also be `embedded' into the Ablowitz--Ladik hierarchy which is usually associated with the evolutionary equations like the discrete nonlinear Schr\"{o}dinger or modified KdV equations.

Next, we would like to point a reader's attention to one of the advantages of the direct approach. If we were trying to derive soliton solutions for the different cases of, say, the general Pohlmeyer sigma-model~\cite{P76} (the so-called~${\rm O}(4)$ version of~\cite{G77} and the ${\rm O}(3,1)$-model of~\cite{V94}) in the framework of the inverse scattering transform,
we would need to elaborate,
actually, two distinct versions of the inverse scattering transform,
due to different analytic structures of the underlying scattering problems.
Even if we compare calculations of this work with ones of~\cite{BG87}, which
are rather universal (in the sense that they can be carried out for different
involution constraints), the method of this paper can be viewed as an useful
alternative. The case is that here all restrictions are formulated in terms
of the constant parameters of the solutions without necessity to discuss the
involution properties of the `intermediate' objects like solutions of the
zero-curvature equations or the corresponding Hilbert problems.

Finally, we would like to note that three models considered in this paper
are far from being the only ones whose solutions can be `extracted' from the
rather simple system \eqref{do:all}--\eqref{unity}. We hope that the studies
presented here can be successfully continued to find other soliton models with
possible physical applications.

\appendix

\section{Derivation of (\ref{do:all})} \label{app:ids}

In this appendix we show how one can calculate derivatives of the functions
defined in \eqref{def:u3v4} and obtain identities from~\eqref{do:all}.

Consider, for example, the function $\v3$. First, we need to calculate the derivative of $\mymat{\G1}$.
The $\o$-dependence of the matrices $\mymat{\A1}$ and $\mymat{\A2}$ defined in~\eqref{def:evol} implies
$ \partial_{\o} \mymat{\A1} = \mymat{\A1} \mymat{\L1}^{-1}$
and $ \partial_{\o} \mymat{\A2} = - \mymat{\A2} \mymat{\L2}^{-1}$
from which one can easily obtain
\begin{equation}\label{appa-1}
 \partial_{\o} \mymat{\A1}\mymat{\A2}
 = \mymat{\A1} \big( \mymat{\L1}^{-1} \mymat{\A2} - \mymat{\A2} \mymat{\L2}^{-1} \big).
\end{equation}
The second equation in \eqref{eq:sy}, after the multiplication by $\mymat{\L1}^{-1}$ from the left and by $\mymat{\L2}^{-1}$ from the right becomes
\begin{equation*}
 \mymat{\A2} \mymat{\L2}^{-1} - \mymat{\L1}^{-1} \mymat{\A2}
 = \mymat{\L1}^{-1} \myket{\ka2} \mybra{\ba2} \mymat{\L2}^{-1}
 = \myket{\ka4} \mybra{\ba4},
\end{equation*}
which, together with \eqref{appa-1}, leads to
\begin{equation*}
 \partial_{\o} \mymat{\A1}\mymat{\A2} = - \mymat{\A1} \myket{\ka4} \mybra{\ba4}.
\end{equation*}
Noting that
$\partial_{\o} \mymat{\A1}\mymat{\A2} =
\partial_{\o} \mymat{\G1}^{-1} =- \mymat{\G1}^{-1}
( \partial_{\o} \mymat{\G1} ) \mymat{\G1}^{-1}$
(we have used the definition \eqref{def:G}) one arrives at
\begin{equation}\label{do:G1}
 \partial_{\o} \mymat{\G1} = \mymat{\G1} \mymat{\A1} \myket{ \ka4} \mybra{ \ba4} \mymat{\G1}.
\end{equation}
Now, using the expression for the derivative of $\mybra{ \ba2}$,
\begin{equation*}
 \partial_{\o}\mybra{ \ba2} = - \mybra{ \ba2} \mymat{\L2}^{-1} = - \mybra{ \ba4}
\end{equation*}
(which follows from the definitions \eqref{def:evol-a} and \eqref{def:baka})
one can derive
\begin{align*}
 \partial_{\o} \v3 & = (\partial_{\o}\mybra{ \ba2}) \mymat{\G1} \myket{ \ka3}
 + \mybra{ \ba2} (\partial_{\o} \mymat{\G1}) \myket{ \ka3}
 = - \mybra{ \ba2} \mymat{\L2}^{-1} \mymat{\G1} \myket{ \ka3}
 + \mybra{ \ba2} \mymat{\G1} \mymat{\A1} \myket{ \ka4}
 \mybra{ \ba4} \mymat{\G1} \myket{ \ka3} \\
 & = - \mybra{ \ba4} \mymat{\G1} \myket{ \ka3} + \mybra{ \ba4} \mymat{\G1} \myket{ \ka3}
 \mybra{ \ba2} \mymat{\G1} \mymat{\A1} \myket{ \ka4} = - \u3 \v7.
\end{align*}

Calculations involving functions $\u3$ and $\u4$ are slightly more complicated. Say, to calculate the derivative $\partial_{\o}\u3$ we, first, rewrite~\eqref{do:G1} in an equivalent form,
\begin{equation*}
 \partial_{\o} \mymat{\G1} = \mymat{\L2}^{-1} \mymat{\G1} - \mymat{\G1} \mymat{\L2}^{-1}
 - \mymat{\G1} \myket{ \ka3} \mybra{ \ba3} \mymat{\G2} \mymat{\A2},
\end{equation*}
which leads to
\begin{align*}
 \partial_{\o} \mymat{\G1}\mymat{\A1}
 & = \mymat{\L2}^{-1} \mymat{\G1} \mymat{\A1} - \mymat{\G1} \mymat{\L2}^{-1} \mymat{\A1}
 - \mymat{\G1} \myket{ \ka3} \mybra{ \ba3} \mymat{\G2} \mymat{\A2} \mymat{\A1} + \mymat{\G1} \mymat{\A1} \mymat{\L1}^{-1}\\
 & = \mymat{\L2}^{-1} \mymat{\G1} \mymat{\A1}
 - \mymat{\G1} \myket{ \ka3} \mybra{ \ba3}
 \mymat{\G2} \mymat{\A2} \mymat{\A1}
 + \mymat{\G1} \myket{ \ka3} \mybra{ \ba3}
 = \mymat{\L2}^{-1} \mymat{\G1} \mymat{\A1} + \mymat{\G1} \myket{ \ka3} \mybra{ \ba3} \mymat{\G2}
\end{align*}
and
\begin{gather*}
 \partial_{\o} \u3= - (\partial_{\o}\mybra{ \ba2}) \mymat{\G1}\mymat{\A1} \myket{ \ka4}
 - \mybra{ \ba2} (\partial_{\o} \mymat{\G1}\mymat{\A1} )
 \myket{ \ka4} = - \mybra{ \ba2} \mymat{\G1} \myket{ \ka3}
 \mybra{ \ba3} \mymat{\G2} \myket{ \ka4} = - \v3 \v8.
\end{gather*}

\section{Proof of (\ref{unity})}\label{app:unity}

Consider the following $(N+2)\times(N+2)$ determinant:
\begin{equation*}
 D = \det\left|
 \begin{matrix}
	1 & \mybra{ \ba2} \mymat{\G1} & 0 \\
	\mymat{\A1} \myket{ \ka4} & \mymat{1}_{N} & \myket{ \ka3} \\
	\langle \ba1 \myket{ \ka4} & \mybra{ \ba1} \mymat{\G2}\mymat{\A2} & 1
 \end{matrix}
 \right|.
\end{equation*}
Calculating this determinant using the Jacobi identity leads to
\begin{equation*}
 D = D_{11}D_{22} - D_{12}D_{21}
\end{equation*}
with
\begin{gather*}
 D_{11} = \det\left|\begin{matrix}
	1 & \mybra{ \ba2} \mymat{\G1} \\
	\mymat{\A1} \myket{ \ka4} & \mymat{1}_{N}
 \end{matrix}\right|
 = 1 - \mybra{ \ba2} \mymat{\G1} \mymat{\A1} \myket{ \ka4} = \u3,\\
 D_{22} = \det\left|
 \begin{matrix}
	\mymat{1}_{N} & \myket{ \ka3} \\
	\mybra{ \ba1} \mymat{\G2}\mymat{\A2} & 1
 \end{matrix}
 \right|
 = 1 - \mybra{ \ba1} \mymat{\G2} \mymat{\A2} \myket{ \ka3} = \u4,\\
 D_{12} = \det\left|
 \begin{matrix}
	\mybra{ \ba2} \mymat{\G1} & 0 \\
	\mymat{1}_{N} & \myket{ \ka3}
 \end{matrix}
 \right|
 = \det\left|
 \begin{matrix}
	0 & -\mybra{ \ba2} \mymat{\G1} \myket{ \ka3} \\
	\mymat{1}_{N} & \myket{ \ka3}
 \end{matrix}
 \right|
 = (-)^{N+1} \v3,\\
 D_{21} =
 \det\left|
 \begin{matrix}
	\mymat{\A1} \myket{ \ka4} & \mymat{1}_{N} \\
	\langle \ba1 \myket{ \ka4} & \mybra{ \ba1} \mymat{\G2}\mymat{\A2}
 \end{matrix}
 \right| =
 \det\left|
 \begin{matrix}
	0 & \mymat{1}_{N} \\
	d & \mybra{ \ba1} \mymat{\G2}\mymat{\A2}
 \end{matrix}
 \right| = (-)^{N} d,
\end{gather*}
where
\begin{equation*}
 d = \langle \ba1 \myket{ \ka4} -	\mybra{ \ba1} \mymat{\G2}\mymat{\A2} \mymat{\A1} \myket{ \ka4}
 =	\mybra{ \ba1}( \mymat{1}_{N} - \mymat{\G2}\mymat{\A2}\mymat{\A1})
	\myket{ \ka4} =	\mybra{ \ba1} \mymat{\G2} \myket{ \ka4} = \v4,
\end{equation*}
or
\begin{equation}
 D = \u3\u4 + \v3\v4.\label{app:Duv}
\end{equation}

On the other hand,
\begin{align}
 D & = \det\left|\begin{matrix}
	1 & \mybra{ \ba2} \mymat{\G1} \\
	\mymat{\A1} \myket{ \ka4} - \myket{ \ka3} \langle \ba2 \myket{ \ka4} &
	\mymat{1}_{N} - \myket{ \ka3} \mybra{ \ba1} \mymat{\G2} \mymat{\A2}
 \end{matrix}\right| =
 \det\left|\begin{matrix}
	1 & \mybra{ \ba2} \mymat{\G1} \\
	\mymat{B} \myket{ \ka4} &
	\mymat{1}_{N} - \myket{ \ka3} \mybra{ \ba1} \mymat{\G2} \mymat{\A2}
 \end{matrix}\right|\nonumber \\
 & = \det\big| 	\mymat{1}_{N}	-	\myket{ \ka3} \mybra{ \ba1} \mymat{\G2} \mymat{\A2}
	-	\mymat{B} \myket{ \ka4} \mybra{ \ba2} \mymat{\G1} \big| = \det\mymat{\G1} \cdot
 \det| \mymat{1}_{N} + \mymat{B} \mymat{\bar{B}}|,\label{app:DBB}
\end{align}
where
\begin{equation*}
	\mymat{B} = \mymat{\A1} - \myket{\ka3} \mybra{\ba1},	\qquad
	\mymat{\bar{B}} = \mymat{\A2} - \myket{\ka4} \mybra{\ba2}.
\end{equation*}
It follows from the Sylvester equations \eqref{eq:sy} that
\begin{equation*}
	\mymat{B} = \mymat{\L2}^{-1} \mymat{\A1} \mymat{\L1},	\qquad
	\mymat{\bar{B}} = \mymat{\L1}^{-1} \mymat{\A2} \mymat{\L2},
\end{equation*}
which leads to
\begin{equation*}
 \det | \mymat{1}_{N} + \mymat{B} \mymat{\bar{B}} | =
 \det | \mymat{1}_{N} + \mymat{\L2}^{-1}\mymat{\A1} \mymat{\A2} \mymat{\L2} |
 = \det | \mymat{1}_{N} + \mymat{\A1} \mymat{\A2} | = ( \det\mymat{\G1} )^{-1},
\end{equation*}
which, together with \eqref{app:DBB}, implies
$D = 1$. Comparing this result with \eqref{app:Duv} one arrives at the identity~\eqref{unity}.

\subsection*{Acknowledgments}

We would like to thank the referees for the constructive comments and suggestions for improvement of this paper.

\pdfbookmark[1]{References}{ref}
\LastPageEnding

\end{document}